# Understanding interlaminar toughening of unidirectional CFRP laminates with carbon nanotube veils

Yunfu Ou[1,2], Carlos González[1,2,*], Juan José Vilatela[1,*]

[1] IMDEA Materials Institute

C/ Eric Kandel 2, 28906 Getafe, Madrid, Spain

[2] E. T. S. de Ingenieros de Caminos, Universidad Politécnica de Madrid, 28040 Madrid, Spain.

**Abstract**

The introduction of nanostructured interlayers is one of the most promising strategies for interlaminar reinforcement in structural composites. In this work, we study the failure mechanism and interlayer microstructure of aerospace-grade structural composites reinforced with thin veils of carbon nanotube produced using an industrialised spinning process. Samples of unidirectional carbon fiber/epoxy matrix composites interleaved with different composition CNT veils were prepared using hot press method and tested for interlaminar fracture toughness (IFT), measured in Mode-I (opening) and Mode-II (in-plane shear), and for interlaminar shear strength (ILSS), evaluated by the short beam shear (SBS) test. The crack propagation mode could be directly determined through fractography analysis by electron microscopy and resin/CNT spatial discrimination by Raman spectroscopy, showing a clear correlation between interlaminar reinforcement and the balance between cohesive/adhesive failure mode at the interlayer region. Composites with full resin infiltration of the CNT veils give a large increase of Mode II IFT (88%) to 1500 J/m$^2$ and a slight enhancement of apparent interlaminar shear strength (6.5%), but a decrease of Mode I IFT (-21%). The results help establish the role of interlayer infiltration, interlaminar crossings and formation of a carbon fiber bridgings, for interlaminar reinforcement with interleaves.

**1. Introduction**

Although fiber reinforced polymer (FRP) composites show many advantages compared to other materials, delamination between reinforcing plies remains a major problem, limiting further implementation of FRP materials. These delaminations are usually formed due to cracks formed under realistic loading conditions, such as impact or fatigue, which can be studied under Mode I and Mode II loading conditions. Once delamination initiates and propagates inside the laminate, the integrity of the whole structure will progressively degrade and eventually result in catastrophic failure. Therefore, the susceptibility to delamination of laminates, often quantified in terms of interlaminar fracture toughness (IFT), is a critical design factor in many applications.

There are ongoing efforts dedicated to developing methods that increase the IFT of composite laminates. 3D reinforcement techniques such as 3D weaving [1], Z-pinning [2] and stitching [3] etc., have proven to be very effective in enhancing IFT, however, these methods often use large filler elements, which can cause a reduction of the in-plane mechanical properties due to lamina fiber damage, stress concentration, and loss of in-plane fiber volume fraction [4, 5]. These negative effects can be avoided by introducing thin toughening layers between lamina, a method known as interleaving. The interleaf materials are usually made of micron-scale toughening particles, thermoplastic films or chopped fibers [6-14]. By generating an interlaminar layer between plies, interleaving may alleviate limitations on plastic zone development in front of the crack tip, thus allowing structures to absorb more fracture energy [10, 11]. Interleaf thickness must be kept small to minimize the weight and thickness of the laminate, and to avoid possible reductions in composite flexural and in-plane properties [12-14].

Nano-scale reinforcement using nanofibers in between the plies of a composite offers the opportunity to reinforce the interlaminar bonding with minimal weight/thickness penalty and can effectively avoid reductions in in-plane mechanical properties. A popular and successful route to improve composite IFT is the use non-woven fabrics of nanostructured fibres, taking advantage of their small thickness, large porosity and inherently high toughness, as well as the possibility to integrate them using existing composite manufacturing methods [15-17]. One family of materials commonly used are electrospun polymer nanofibers [15-23]. Appropriately chose, the thermoplastic can mix with the composite polymer matrix and act as a toughening agent [24-26].

Several approaches by integrating CNTs or graphene into FRP structural composites for improving the interlaminar properties have been studied [27-33]. Dispersing CNTs or graphene throughout the entire composite matrix has been a popular route to seek interlaminar reinforcement, but the dramatic increase in viscosity upon addition of nanoparticles represents a fundamental processing hurdle, often leading to incomplete textile impregnation [34]. Spraying of CNTs or transfer CNT forests directly on fibre preforms or prepreg also draw much attention since it overcomes above limitations such as increased resin viscosity by CNTs or uneven distribution and filtering during resin infusion processes, although scalabitiy remains a major issue. Another popular route has consisted in the growth of CNTs on the surface of CF. After improvements in the synthesis method to avoid degradation of the CF by the metallic catalyst, interlaminar properties have been improved while preserving longitudinal composite properties [35].

Recently, much attention has been paid to the use of thin veils of carbon nanotubes (CNT) as interleaves in laminated composites [36-40]. These veils can be intercalated between the primary reinforcing fiber layers before infusion, so there is no need to disperse them into the resin. Therefore, there are no disadvantages such as increased viscosity from dispersing nanofillers in resin and uneven distribution or segregation from filtration of nanofillers [32, 33]. Compared with electrospun thermoplastic interleaves, CNT fiber veils have order-of-magnitude higher surface area and fracture energy. They are also thermally and chemically stable, and are already produced industrially [41], thus making them a promising interleaf material for widespread implementation in structural composites. CNT veils can be made extremely thin (<50 nm), and with sufficiently high longitudinal Young's modulus to preserve interlaminar rigidity, if properly integrated.

Although the use of CNT fiber veils offers many potential benefits compared to traditional toughening methods, the research on composites enhanced with CNT fiber veils is limited and often contradictory. Nguyen et al. [40] introduced aligned CNT sheet into unidirectional CFRP system. Mode II interlaminar fracture toughness was found to be significantly improved (~70%), while Mode I fracture toughness slightly decreased. This reduction was attributed to partial impregnation of epoxy into the thick CNT film and the sheet forming an interlaminar layer which acts as a barrier for micro-scale fiber bridging. Nistal et al. [37]

interleaved an ethylenediamine-functionalised multilayer CNT web (0.2 g/m$^{-2}$) between CF plies, and reported a 13% enhancement in the Mode I IFT. Di Leonardo et al. [38] introduced pristine and functionalized carbon nanotube (CNT) webs at the interlaminar region of two different thin-ply CFRP systems (NTPT 402 prepreg, and TeXtreme/SE84LV). Experimental results showed that the Mode I IFT for the NTPT 402 system was substantially lower while there was no reduction in the steady state value of the Mode I IFT when SE84LV resin was used. Our recent work [36] demonstrated an improvement of 60% in Mode-I fracture toughness of a model system of woven CF interleaved with low-density "fluffy" CNT veils, and a 28.4% decrease when CNT veils of lower permeability were integrated. These emerging results demonstrated the potential of reinforcing the interlaminar region with different kinds of CNT veils/webs (as-spun, functionalized or densified), but remain inconclusive and of limited generality because some fundamental aspects of CNT veil/web toughening remain unclear. Often overlook, for example, is the degree of resin infiltration into the CNT veils. It is hard to control and difficult to evaluate, but resin infiltration is expected to depend on resin properties as well as on veil morphology, thickness and surface chemistry (functionalization). Overall, a systematic interlaminar test campaign and a thorough understanding of the fracture and toughening mechanism is clearly needed in order to link the structure of CNT fiber veils to their effects in IFT.

In the present work, we demonstrate a facile and scalable method to integrate CNTs into unidirectional laminate composite. CNT fiber fabric veils with highly porous structure are directly affixed to the surface of carbon fiber prepreg without any treatment, continuously, and directly as they are synthesized in chemical vapour deposition reactor. Samples with different interleaf composition were fabricated utilizing hot press method, and subjected to Double cantilever beam (DCB), End-notched Flexure (ENF) 3-point bend as well as Short beam shear (SBS) testing to evaluate the interlaminar fracture toughness (Mode I and Mode II IFT) and interlaminar shear strength (ILSS), respectively. Extensive fractography by Raman spectroscopy and electron microscopy was enabled determination of the micromechanisms and fracture behaviors of virgin and nanofiber interleaved composite laminates, and the predominant role of resin infiltration into the nanostructured interlayer.

## 2. Materials and methods

*2.1 Materials*

CNT veils were synthesized by the direct spinning process from the gas-phase during growth of CNTs by floating catalyst chemical vapor deposition (FCCVD[42]), as shown in **Figure 1a**, utilizing ferrocene (1.2%) as iron catalyst, thiophene (1.2%) and Toluene (97.6%) as sulfur catalyst promoter and carbon source, respectively. The reaction was carried out in the hydrogen atmosphere at 1300 °C with precursor feed rate of 3.7 mL h$^{-1}$ and a winding rate of 20 m min$^{-1}$. The CNT fibers were continuously drawn out at the exit of the furnace and directly deposited onto the surface of a piece of unidirectional carbon fiber prepreg (unidirectional UD HexPly® AS4/8552 prepreg with 34% resin content and 194g/m$^2$ areal weight, as shown in **Figure 1b**), which was wrapped on a rotating drum (D = 10 mm). CNT fiber winding was carried out following the carbon fiber direction with a deposition time of 20 min, which gave an areal density of 0.8 g/m$^2$. A high-resolution micrograph of the as-produced CNT fiber veil is presented in **Figure 1c** showing an ultrahigh porosity.

*2.2 Testing standard, laminate manufacturing and specimen preparation*

Interlaminar fracture toughness was measured by using DCB and ENF specimens for mode I and II crack propagation, in accordance with ASTM standard D5528-01 [43] and ASTM standard D7905 [44], respectively. Three-point bending tests on SBS specimens were also conducted to evaluate the interlaminar shear strength, based on the ASTM standard D2344 [45].

2 panels of 20 plies unidirectional carbon fiber prepreg (280 mm × 280 mm) following the [0°]$_{20}$ lay-up with a nominal thickness of 3.65 mm were prepared by compression moulding. The dimensions allowed to extract 10 DCB, 10 ENF and 10 SBS specimens (five are with CNT and five without) as reported in **Figure 2d**. In both of the two panels, the mid-plane ply (11$^{th}$ in the lay-up) was covered with fluffy as-produced CNT veils (see **Figure 1b**). In order to guarantee a good resin impregnation of the CNT veils, the 10$^{th}$ prepreg ply of one panel was stuck with a thin 8552 resin film layer (25 μm), facing with the CNT veils on the 11$^{th}$ ply. A 25 μm-thick teflon film was inserted at the mid-plane as a crack starter in the mode I fracture tests. An additional panel was prepared by interleaving only with the 8552 resin film in order to ascertain its effect without CNT veil reinforcement. Hot-Plate Press machine (LabPro 400, Fontijne Presses) was then used to consolidate laminate panel from pre-impregnated sheets of fiber-reinforced composites by simultaneous application of pressure (up to 7 bar) and heat (up to 180 ºC) according to the recommendations of the prepreg supplier. The panels were submitted to ultrasound inspection to ensure they were free of manufacturing defects, voids or delaminations. DCB specimens were extracted from laminates utilizing a water-cooled milling machine. The lateral side of each specimen was then sprayed with white primer paint and the 120 mm after the pre-crack marked in 1 mm increments to track crack growth throughout the test. It is worth noting that ENF specimens were directly extracted from previously tested DCB specimens (**Figure 2e**), which is quite material-saving and creates a more real pre-crack with a small tip radius for ENF test. SBS specimens were precisely cut utilizing a diamond-wire cut machine.

*2.3 Characterizations*

Ultrasound inspection was performed to determine the quality of the panel and to localize possible manufacturing defects. FIB-FEGSEM microscope (Helios NanoLab 600i FEI) were used to investigate the morphology of CNT fiber veils and the fracture surfaces of the laminates. An optical microscope (OLYMPUS BX51) was used to visualize the distribution of CNT veils in the cross section of the composite laminates and then to determine the thicknesses of the CNT layer. Raman spectroscopy (Renishaw PLC) was performed to assess the impregnation level of the resin inside the CNT veils. An excitation wavelength of 532 nm from a tunable YAG laser was focused on the sample using laser power in the range of 1-5 mW so as to avoid sample heating.

**3. Results and Discussion**

*3.1 Laminate examination*

The composite panels were 280 × 280 mm, and purposely designed to comprise an interleaved area of 140 × 280 mm and the remaining part free of interleaf to be used as control specimen. Four types of samples were produced in order to ensure a complete study. Baseline panels were manufactured using AS4/8552 prepregs as well as laminates interleaved with an additional 8552 epoxy film (resin interleaf-AS4/8552). Similarly, a first sample type consists of as-received AS4/8552 prepregs interleaved with CNT veils (CNT interleaf-AS4/8552), whereas a second one included the introduction of an additional 8552 epoxy film (CNT/resin interleaf-AS4/8552). Reference specimens were extracted from each same panel to avoid any variations between different batches, although, in general, the fabrication process was optimized to avoid variations between different panels.

**Figure 3** shows an example of a final CF/CNT veils/epoxy hybrid composite produced using the aforementioned hot-press compression molding. A plot of the attenuation signal obtained from the ultrasound inspections (**Figure 3a**) showed that there exist no obvious defects/voids inside the laminate, without significant thickness variations proved by a homogeneous attenuation distribution. The cross-sectional of the laminates is shown in (**Figure 3b**), which was obtain by means of optical microscope and further confirmed almost void-free laminate production showing that the CNT veil was fully integrated in the composite with a thickness of $10.0 \pm 1.0$ microns.

*3.2 Mode I IFT*

Two different data-reduction of fracture toughness values can be used to describe the fracture behaviour of the laminates, namely initiation and propagation toughness ($G_{Ic,ini}$ and $G_{IC,prop}$), respectively. $G_{Ic,ini}$ is determined by the 5%/max point in the force-displacement curve after precracking, while the $G_{IC,prop}$ is the value where the energy release rate achieves a constant state and the crack propagates in a self-similar way independent on the crack length [43]. The $G$ value of interest for this experiment was $G_{IC,prop}$, since it is more realistic in practice for determining the likelihood of the crack to continue to propagate.

**Figure 4a** and **b** displayed the representative load-displacement curves for mode I interlaminar fracture tests. All the test exhibited a stable and smooth crack propagation. The corresponding R-curves of the different samples, respectively were obtained following the data-reduction methods aforementioned. For the control baseline samples (AS4/8552 prepregs without interleaf), a relatively large increase in $G_I$ with the increasing crack length is observed which is predominately attributed to carbon fiber bridging. Conversely, the samples interleaved with CNT veils showed a relatively flat R-curve behaviour which was an indicative of fiber bridging suppression. More specifically, the $G_{IC,prop}$ values decreased as much as 71% (compared with control samples) when directly interleaving CNT veils. Such significant knock-down in Mode I fracture toughness can be ascribed to the poor resin infiltration of the prepreg resin bleeding into the CNT veils. This fact was confirmed by the SEM picture of the corresponding fracture surface as shown in **Figure 4e** and by the Raman analysis (see **Section 3.5.1**). It seems that resin bleeding from the AS4/8552 prepreg was not enough to produce good infiltration into the CNT veil. To alleviate such effect, 8552 resin films were intercalated together with the as-produced CNT veil in the middle layer of the manufactured panels. Interestingly, it was found the intercalation of such resin film improved significantly the resin infiltration of the CNT veils (see **Section 3.5**), thus enhancing the Mode I mechanical performance. Compared with samples only interleaved with CNT veils, $G_{IC,prop}$ values were improved as much as 170%. It is worth noting that the $G_{Ic,ini}$ value of CNT veil/resin film laminates maintained at the same level as the control sample without interlaminar reinforcement while the $G_{IC,prop}$ value was 21% smaller. In addition, 8552 epoxy films were also intercalated in the baseline AS4/8552 laminates to isolate its influence on the fracture behavior. In this case, the Mode I fracture toughness of the only resin-interleaved specimen showed no significant statistic changes compared with that of control samples (see **Figure 4c**). Therefore, the results suggest that CNT veils require a minimum of resin impregnation to transfer some load between crack wakes.

To summarize, the inclusion of CNT veils seems to provide a preferential path for crack propagation hampering carbon fiber bridging, which is known to be one of the major contributions to mode I interlaminar fracture toughness. The carbon fiber bridging observed in CNT interleaved samples was negligible and, thus, any contribution to the fracture toughness from this source is likely to be minimal. Although both CNT interleaved laminates exhibited a significant decrease of the fracture toughness, and this fact was attributed to the absence of fiber bridging, the reduction obtained in the case of the CNT film interleaved laminates with the supplementary 8552 resin film was less pronounced indicating some rudimentarily CNT stress transfer.

Moreover, it should be emphasized that that toughening of resins with dispersed CNT is often perceived to be the result of CNTs being oriented orthogonal to the fracture plane acting as a kind of 'nano-stitches'[46]. However, the CNTs used in this work are predominantly oriented parallel to the plane of the veil, therefore, the loading of the CNTs during mode I crack propagation was not optimal. As the crack is initiated in the CNT modified interlayer growing in mode I, the CNT veils are loaded transversally and subjected to out-of-plane peeling forces. The CNTs in the fracture surface did not exhibit neither stretching nor loading alignment (see **Figure 4e and f**), suggesting that most of the CNTs seem to have been peeled off from the matrix resin without significant deformation. The introduction of the CNT veil interleaf gave rise to coupled detrimental effects: a low energy-dissipation path tracking the crack propagation and at the same time limiting the extension of carbon fiber bridging as the main contributor to the interlaminar material toughness. Those mechanisms produced the substantial reduction of the carbon fiber bridging as compared with the baseline laminates.

*3.3 Mode II IFT*

For the Mode II ENF tests, specimens were extracted from the previously tested Mode I coupons according to the standard requirements. Representative load-displacement curves and the corresponding Mode II IFT ($G_{IIC}$) for each laminate configuration (baseline, resin interleaf-AS4/8552, CNT interleaf-AS4/8552 and CNT/resin interleaf-AS4/8552) are given in **Figure 5**. It was found that resin interleaved samples also showed the similar mechanical responses with control specimens under mode II loading condition as it was observed in mode I tests. However, only the introduction of both CNT veil and resin film resulted in a higher maximum load at the onset of fracture and thus in a higher $G_{IIC}$ value. In this case, an 88% enhancement was obtained as compared with baseline samples without interleaves. It should be pointed out that when using the as-produced CNT veils alone without 8552 epoxy resin film (CNT interleaf-AS4/8552), only a slightly improvement of $G_{IIC}$ of 12% is measured. Again, the AS4/8552 epoxy prepreg resin bleeding was not enough to fully impregnate the CNT interleaf, resulting in a minor improvement. By scrutinizing the fracture surfaces, three different failure mechanisms can be distinguished which are schematically shown in **Figure 5 (f, g and h)**. The failure mechanism for the baseline is shear dominated interfacial failure at the fiber/matrix interface, which was confirmed by SEM image (**Figure 5c**). The $G_{IIC}$ in this case strongly depends on the bond strength between carbon fiber and resin. The introduction of a CNT veil in the mid-plane of the laminates resulted in the formation of an interlaminar region with a thickness of around 10 μm. When CNT veil is poorly infiltrated by the resin, a preferential crack path is created in this layer, resulting in a cohesive dominated failure (**Figure 5g**). The high-resolution micrograph of the fracture surface presented in **Figure 5d** showed that the crack exclusively propagates through the CNT veil interlayer. This phenomenon was also observed in our previous work [47], in which commercial CNT film was used. It should be noted, in this case, that the CNTs were highly aligned with the mode II shear loads at the crack tip so some effective bridging could occur resulting in a 12% improvement of $G_{IIC}$ as previously mentioned. When CNT veil is well infiltrated, crack tip is deflected away from the CNT toughened interlayer into the interface of CF ply and toughened interlayer (adhesive failure), resulting in interlaminar crossing and effective CNT crack bridging (**Figure 5e and h**). In this scenario, the failure mechanism not only includes the debonding of carbon fiber from CNT modified matrix, but also interlaminar crossing and bridging at points where the crack passes from one side of the interlayer to the other, significantly enhancing the Mode II IFT.

*3.4 Interlaminar shear strength*

The small size of the ILSS test specimen combined with the ease of testing make the test method more attractive than other shear tests [48]. The short span of the beam reduces the bending effects loading the mid-plane under shear dominated forces. **Figure 6a** displays the representative load-displacement curves of each composite configuration. They are almost linear and elastic, except at the peak load. The maximum force attained during the tests was used to determine the interlaminar shear strength of the laminate according to the standard assuming the maximum shear stress produces cracking at the mid-plane position of the laminate. The ILSS statistical results shown in **Figure 6b** are average values of at least 3 individually tested specimens for each laminate configuration. In all the cases, single/multiple cracks propagating parallel to the span direction were observed emanating from one of the supports of the beam and not necessarily located in the mid-plane of the laminate. Thus, the ILSS strength obtained via **Formula (1)** based on ASTM standard D2344 [45] should be treated as an apparent value of the shear strength. Not surprisingly, the statistic results showed that interleaving thin resin film exerts negligible influence on the interlaminar shear properties of control laminate.

$$\text{ILSS} = \frac{3}{4} \frac{F_{max}}{wt} \qquad (1)$$

The apparent ILSS of unidirectional laminate samples decreased from 101.1±2.1 MPa to 93.1±3.7 MPa when interleaved with only CNT veils. Such relative decrease was around -7.9% (see **Figure 6b**). When looking into their specific failure modes as shown in **Figure 6c** and **d**, we found multiple shear cracks running along the fiber direction of the control sample included the mid-plane one (**Figure 6c**). As a contrast with baseline samples, only one crack was observed just in the middle plane of the CNT veils interleaved sample (**Figure 6d**). When interleaving with resin films and as-produced CNT veils in the mid-plane of the specimen, the apparent ILSS increased from 101.1±2.1 MPa to 107.1±2.4 MPa, enhanced then by 6.5%. Cross-sectional micrograph in **Figure 6e** showed that the multiple cracks were developed near the mid-plane, meanwhile this interface itself still kept intact, indicating that CNT toughened region bears higher ILSS than the control interlayer.

It should be emphasized that the ILSS determination using Formula (1) is on the condition that crack was initiated in the middle plane. For samples interleaved with both CNT veil and resin film, cracks were initiated and propagated near the middle layer, while the middle layer itself still kept intact. In this case, Formula (1) is not applicable. Based on linear and elastic calculations and some basic assumptions as shown in **Figure 7a,** following inequality was finally obtained:

$$\tau_{max}^{CNTR} \geq \frac{\tau_{max}^{BL}}{1-\alpha^2} \qquad (2)$$

Where, $\alpha$ is the ratio of the distance between the crack initiation plane and the middle plane to the half thickness of the laminate, $\tau_{max}^{CNTR}$ and $\tau_{max}^{BL}$ are ILSS of CNT veil/resin film interleaved specimen and baseline, respectively.

**Formula (2)** provides the lower limit of ILSS for CNT veil/resin film interleaved specimens. In this paper, $\tau_{max}^{BL}$ is determined as 101.1MPa (**Figure 6b**). The distance between the crack initiation plane and the middle plane is around 300 μm as shown in **Figure 7b**, and the thickness of laminate (t) is around 3.65 mm, so $\alpha$ is calculated as 0.16. Thus, $\tau_{max}^{CNTR} \geq$ 103.8 MPa.

*3.5 Discussion*

*3.5.1 Resin film and veil impregnation*

It was found that interleaving thin 8552 resin film exerted negligible influence on the interlaminar properties of 8552/AS4 composite laminate. $G_{IC}$, $G_{IIC}$ and ILSS were nearly unchanged after film integration. **Figure 8** showed the cross-sectional image of resin-interleaved laminate, in which no significant difference was found between the mid-plane region (interleaved with resin film) and other interlaminar regions. The film in this case is very thin (~25 microns) so resin can be easily distributed during the cure cycle.

Interestingly, locally interleaving with thin resin films together with the CNT veils can significantly improve its resin infiltration. This fact was confirmed subsequently by the SEM high resolution images of the corresponding fracture surfaces as shown in **Figure 4 (e)** and **(f)**. The composition of the different regions could be readily determined from their Raman spectra taking advantage of the strong Raman intensity and distinctive features of the CNT veils compared with the carbon fiber and the polymer resin. Considering together the low D/G band ratio and the sharp G band line-shape as displayed in **Figure 9a**, the fracture surface of only CNT veil interleaved sample can be easily recognized as dry CNT-dominated regions in the fracture surface. In comparison, a prominent fluorescence effect was recorded during the Raman analysis on the fracture surfaces of CNT veil/resin film interleaved specimens which confirmed the adequate infiltration of resin into the CNT veils.

*3.5.2 Toughening mechanisms*
It is common that laminate composites incorporated with same interleaf materials perform quite differently in Mode I and Mode II loading conditions [15, 16]. Some material properties may exert a positive or a negative influence on each mode of interlaminar failure. Both kinds of failures are complicated, with Mode I case being proveiled by peel stress and Mode II being dominated by shear stress, and it is really a challenge to figure out which material properties or their combinations are most likely to affect crack initiation and propagation behaviors, thus changing the final fracture toughness.

The remarkable enhancement of Mode II IFT of CNT veil interleaved laminates (well-infiltrated) can be attributed to the ideal loading case for the CNT veils. The CNT veils used in this paper are primarily oriented parallel to the fracture plane, which is designed to carry load in the plane, rather than out of plane. Under Mode II loading condition, stresses are effectively transferred to the veil by membrane mechanisms which enhance the interlaminar toughness $G_{IIC}$ values as it was demonstrated in the previous paragraphs. On contrast, under Mode I loading condition CNTs are mostly peeled out from the assembly, which can not take full advantage of their real strength. Our previous work [36] demonstrated improvements of 60% in Mode-I fracture toughness after interleaving similar CNT veils into woven carbon fabrics. It was this woven structure that reshaped the CNT veil configuration after compression, which was prove to be favor of stress transferring under Mode I loading state.

Toughening effect, to a considerable degree, depends on crack behavior [36] and the ability of the material for crack bridging. As shown in **Figure 10**, the Mode I fracture surface of CNT veil toughened samples was quite smooth (**Figure 10a, up**) and CNTs were universally found in both arms of fracture samples (**Figure 10b, right**). Thus, it can be concluded that the crack progresses almost exclusively along the CNT-rich interlayer, corresponding to a cohesive failure (**Figure 10c, up**). Under the Mode II load condition, a "banded" structure in the specimen loading direction on the fracture surface was found (**Figure 10a, bottom**). Closed examination under SEM showed that adhesive failure was dominated in this case (**Figure 10b, left**). It should be noted that those adhesive failure regions were not uniformly distributed along one side of CNT veils, but alternately distributed in the both sides, which leads to a lot of interlaminar crossings (**Figure 10c, bottom**). In fact, the black bands in bottom picture of **Figure 10a** are originated from those interlaminar crossings. Moreover, the interfaces between CF and resin are modified by the CNTs (**Figure 10b, left**), which could have a huge contribution to the Mode II IFT.

The toughening effect of CNT veils also depends on the architectures of the host fabrics. In unidirectional laminate, the fabric ply is quite flat and smooth, hence the majority of CNTs are parallel with the delamination plane of the DCB specimen. As the crack is driven by Mode I (opening) loading condition, the CNTs are loaded transversally and subjected to peeling forces. It is worth noting that similar phenomena were also observed when using electrospun nanofiber veils in UD composites[15, 49]. Daelemans et al. [15], for example, reported that the presence of the electrospun nanofibrous veils between UD plies blocked the formation of a carbon fibre bridging zone and most of the nanofibres seem to have been peeled off from the matrix resin without much deformation during the crack propagation under Mode I loading condition. Interestingly, they also found if a zone of carbon fiber bridging was able to develop before the delamination encountered the nanofiber modified region, the Mode I fracture toughness increased notably in this nanofiber modified region. In our previous work [36], when 5H satin woven fabric was utilized, Mode I interlaminar fracture toughness was enhanced as much as 60%. In this scenario, the woven architecture of the fabrics probably resulted in tiny regions where the local crack growth was not purely Mode I or where the crack growth direction was not perfectly parallel with the plane of the interlayer, triggering a better load transfer to the CNTs and more fiber-bridgings, resulting in a significant enhancement of fracture energy consumption.

*3.5.3 Summary*

The novelty of this work can be rooted in three sides. One is Mode II toughening factor ($\eta_{II}$), which can be defined as the improvement of Mode II interlaminar fracture toughness ($\Delta G_{IIC}/G_{IIC}$) divided by relative-to-ply interleaf thickness. In this work, we demonstrated an 88% improvement of $G_{IIC}$ by interleaving a layer of 10um thick CNT veils (veil/ply ratio ~5.5%). The calculated $\eta_{II}$ value (88/5.5 ≈15.8) ranked top among other interleaf materials listed in the **Table 1**. It should also highlight that weight/thickness penalty can also be minimized when thinner interleaved are used, as well as the possible influence of other mechanical performance (i.e bending, in-plane shear, etc.)

Another merit of this paper is that we demonstrate a facile and scalable manufacturing method in which CNT veil is directly wound onto the prepreg surface. Most of reported interleaving materials/methods are either complicated to make/execute, or difficult to be realized in large scale. In this work, we try to avoid any post-processing to the materials and scale-limited steps, even though we know some of them (i.e. functionalization [37]) would exert a positive influence on the fracture behaviors. In addition, the integration of pre-assembled CNT veils without recourse to solvents is also potentially compatible with standard composite manufacturing methods and essentially analogous to the introduction of electrospun thermoplastic veils already utilized industrially [17]. Recently, this method has been successful applied in a joint research work with a company. We not only demonstrated a significant improvement of interlaminar fracture toughness, but also proved the scalability of this technique. Related results will be published soon.

Last but not least, we emphasize that our aim is not only to demonstrate in increase in toughness, but also to provide a rationale for the observed results. The new Raman fractography method and systemic test campaigns of interlaminar properties, are examples of developments that improve general understanding of interlaminar reinforcement with CNT veils.

## 4. Concluding remarks

The objective of this work is to investigate the possible effect of CNT veil interleaves on interlaminar properties of unidirectional CFRP laminate composite. A facile and scalable hot press protocol was successfully built to make CNT/CF/epoxy hybrids, by which void-free laminates can be manufactured. The IFT under Mode I and Mode II loading cases and ILSS are investigated and discussed comparatively, followed by a systematic analysis of failure and toughening mechanisms. The following conclusions can be reached:

Integration of thin 8552 resin film exerted negligible influence on the interlaminar properties of AS4/8552 composite laminate. $G_{IC}$, $G_{IIC}$ and ILSS were proved to be unchanged after film interleaving.

Only interleaving as-produced fluffy CNT veils into unidirectional CFRP system led to an increase of Mode II IFT (12%) but a significant decrease of Mode I IFT (71%), as well as a slight degradation of ILSS (-7.9%). It is common for a same interleaving system varies dramatically in Mode I and Mode II toughening effects because of totally different stress state under respective mode, with Mode I being dominated by peel forces and Mode II being dominated by shear forces.

Locally adding resin film near CNT veils can substantially improve the infiltration condition and thus interlaminar properties. Compared with samples only interleaved CNT veil, $G_{IC,prop}$ was improved as much as 170% after supplementing a thin resin film. Compared with control samples, the Mode II IFT and apparent ILSS were increased by 88% and 6.5%, respectively. The depreciation of $G_{IC,prop}$ value probably was due to the in-plane orientation of the CNT veil within the ply that caused the low peeling resistance.

**Acknowledgements**

The authors are grateful for generous financial support provided by the European Union Seventh Framework Program under grant agreement 678565 (ERC-STEM), the MINECO (RyC-2014-15115, MAT2015-64167-C2-1-R), the ESTENEA funded by AIRBUS Operations S.L. and CDTI (CIEN 2014 program), SORCERER (funded by the European Union, Clean Sky Joint Undertaking 2, Horizon 2020 under Grant Agreement Number 738085) and the Cost Action CA15107 (MultiComp). Yunfu Ou would like to appreciate the financial support from the China Scholarship Council (grant number 201606130061) and technical assistance from Vanesa Martínez and José Luis Jiménez in Imdea Materials.

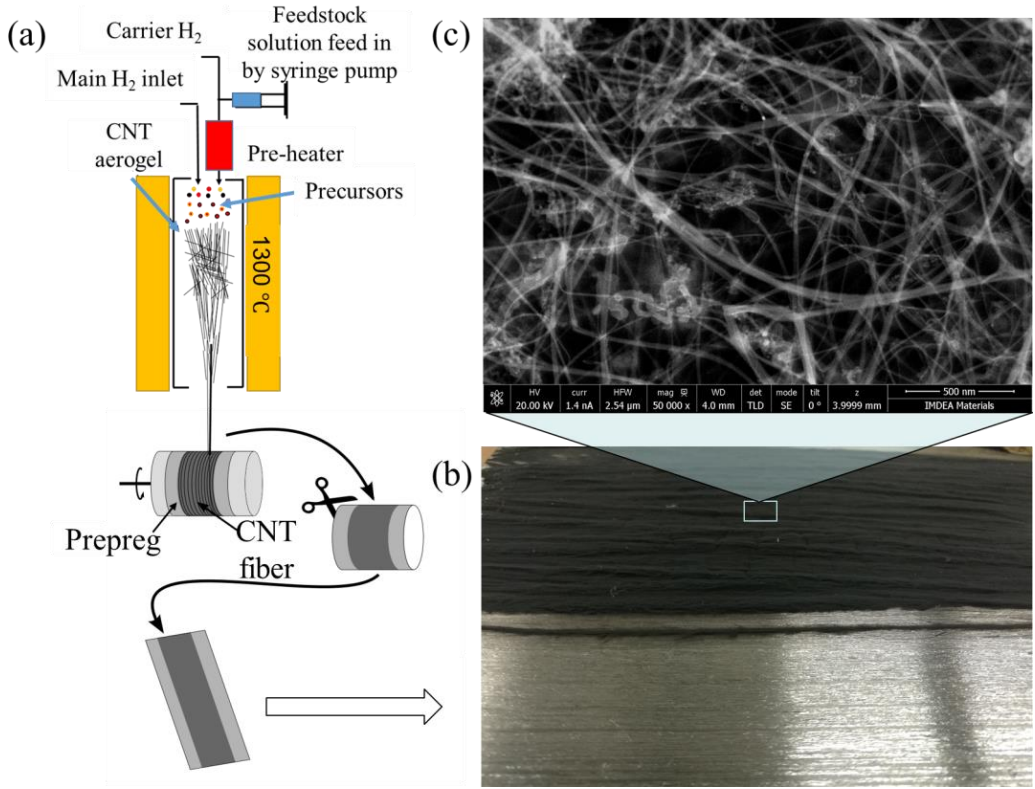

Fig. 1 (a) Schematic of direct spinning of CNT fibers from the gas phase by FFCVD; (b) unidirectional carbon AS4/8552 prepreg with fluffy CNT veils deposited on the surface and (c) high resolution SEM picture of as-produced CNT fiber veil

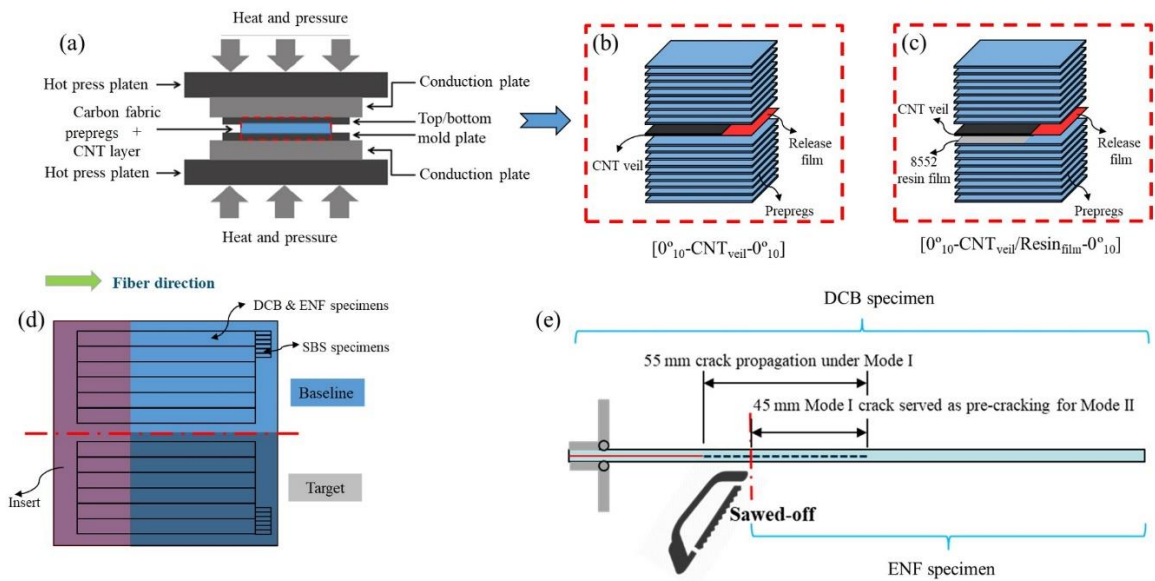

Fig. 2 (a) schematic of hot press method; (b) and (c) lay-ups of prepregs in which 2 different interleaves were integrated; (d) the design of the panel from which different specimens are cut and (e) a new conception for preparation of mode II sample.

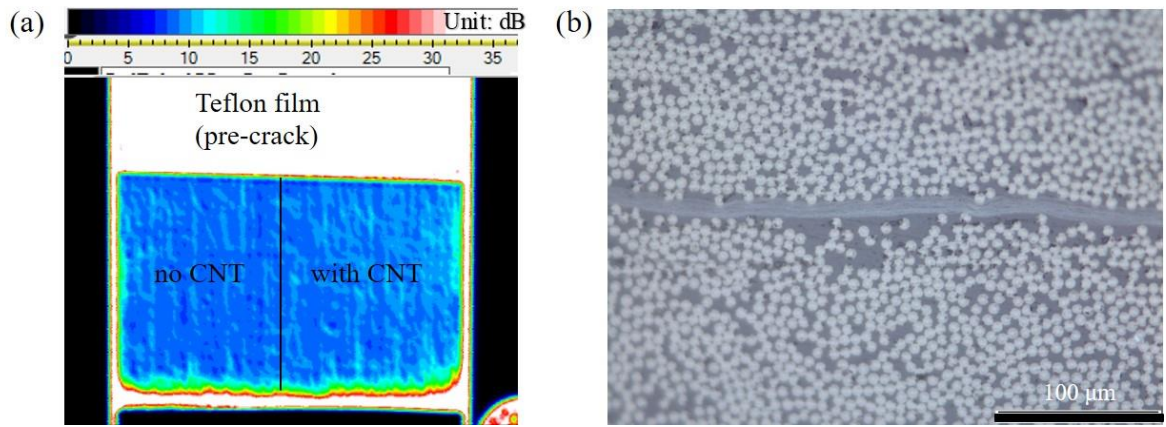

Fig. 3 (a) Attenuation plot of the laminates containing CNT veils obtained by ultrasound inspection (b) cross-section of the laminate obtained by optical microscopy.

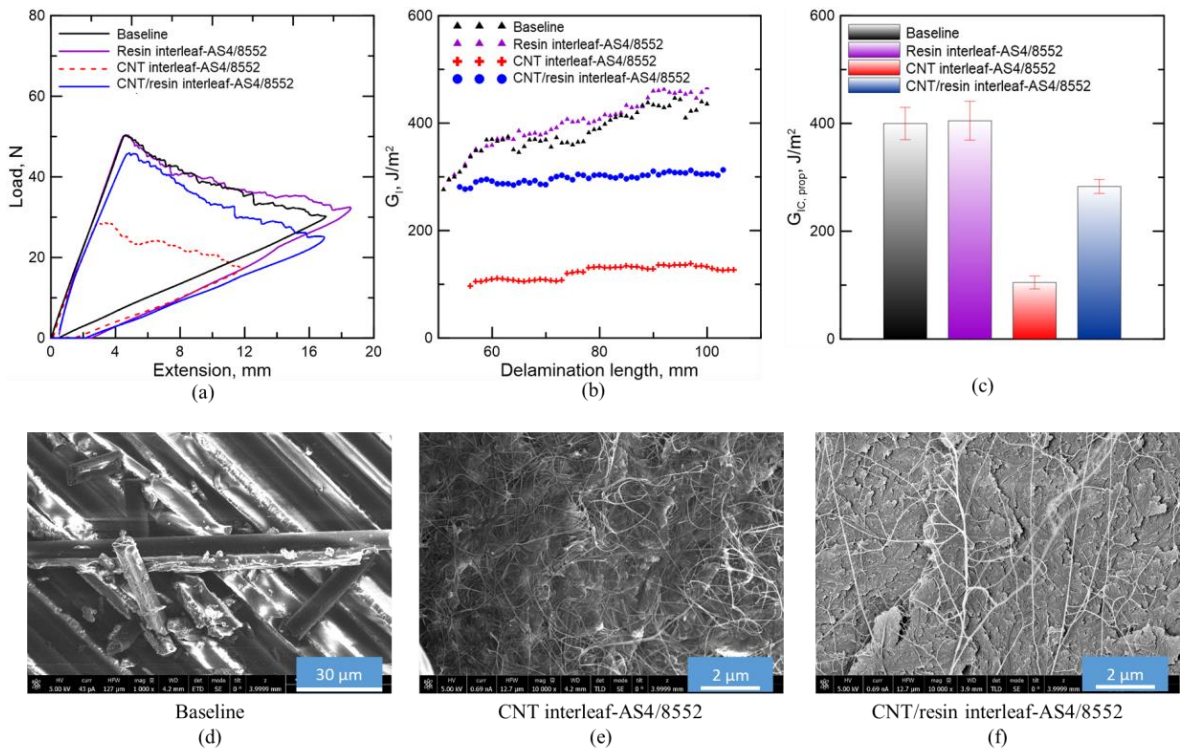

Fig. 4 Representative load-displacement curves (a) and the corresponding R-curves (b) of specimens with and without CNT veils in Mode I loading case; (c) comparison of Mode I IFT for CNT-interleaved composites and baseline; (d) (e) and (f) are the corresponding SEM micrographs of the fracture surfaces.

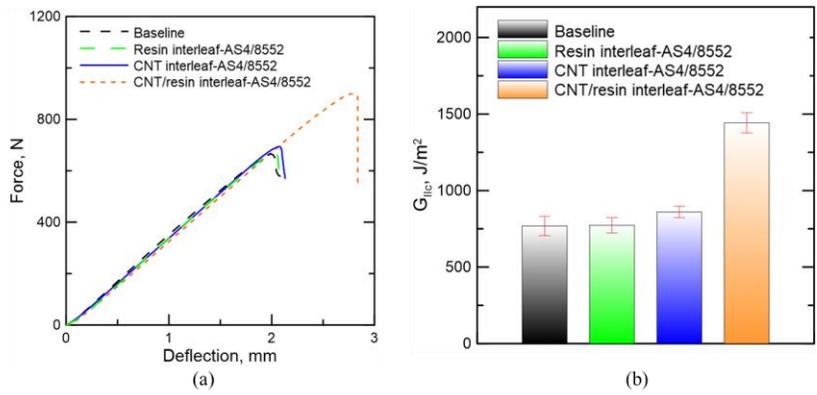
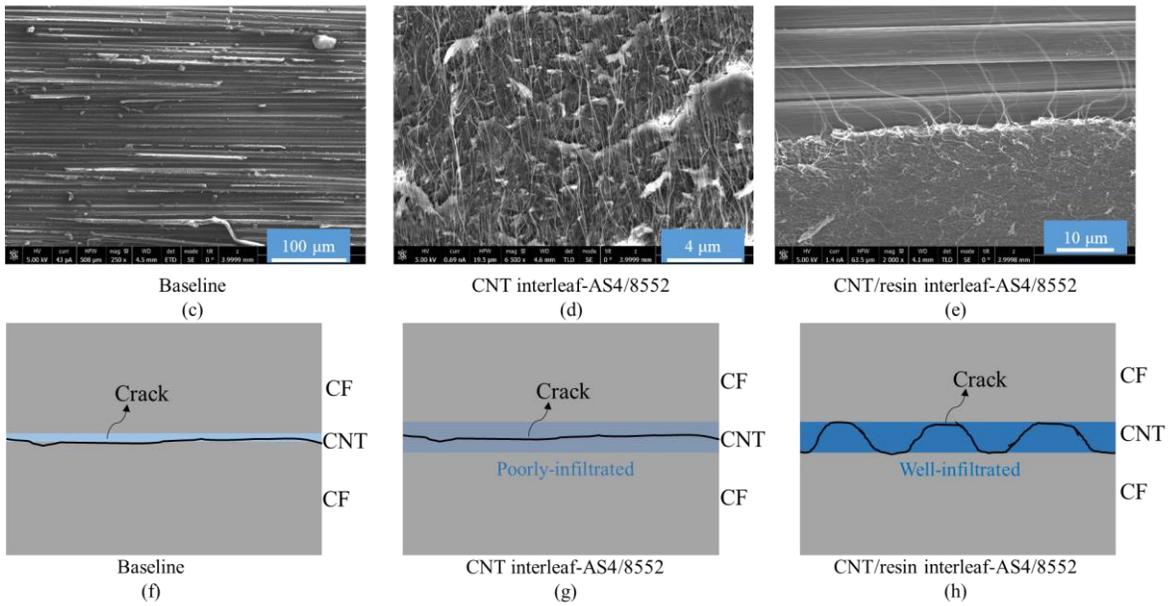

Fig. 5 Representative Load-displacement curves of specimens with and without CNT veils in Mode II tests; (b) comparison of Mode II IFT for CNT-interleaved composites and baseline; (c), (d) and (e) are the corresponding fracture surfaces under SEM; (f), (g) and (h) sketches of possible fracture mechanisms of the three specimens analyzed.

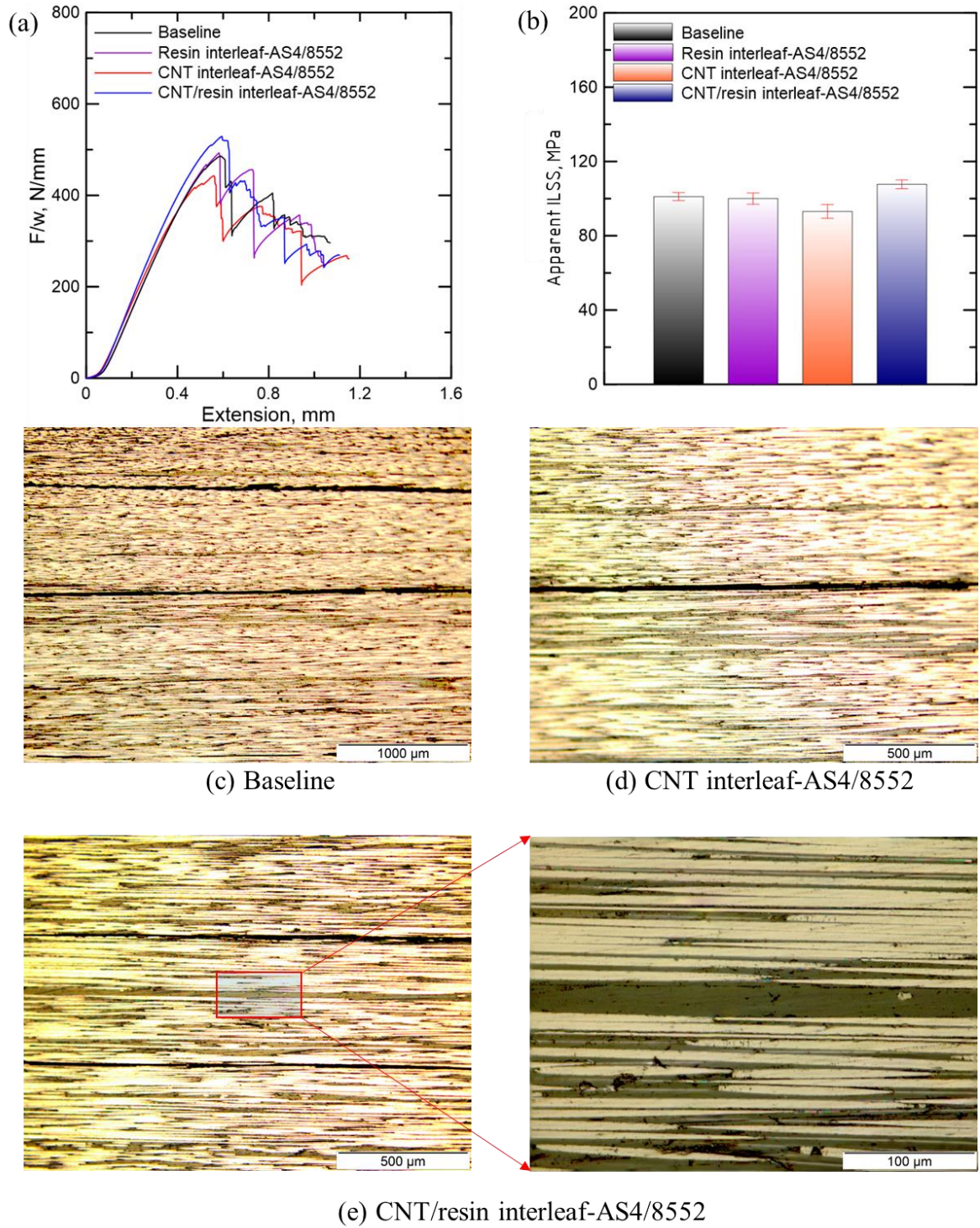

Fig. 6 Representative Load-extension curves of specimens with and without CNT veils in short beam shear test; (b) comparison of apparent ILSS for CNT-interleaved composites and baseline; (c) (d) and (e) are the corresponding cross-sectional images under optical microscope.

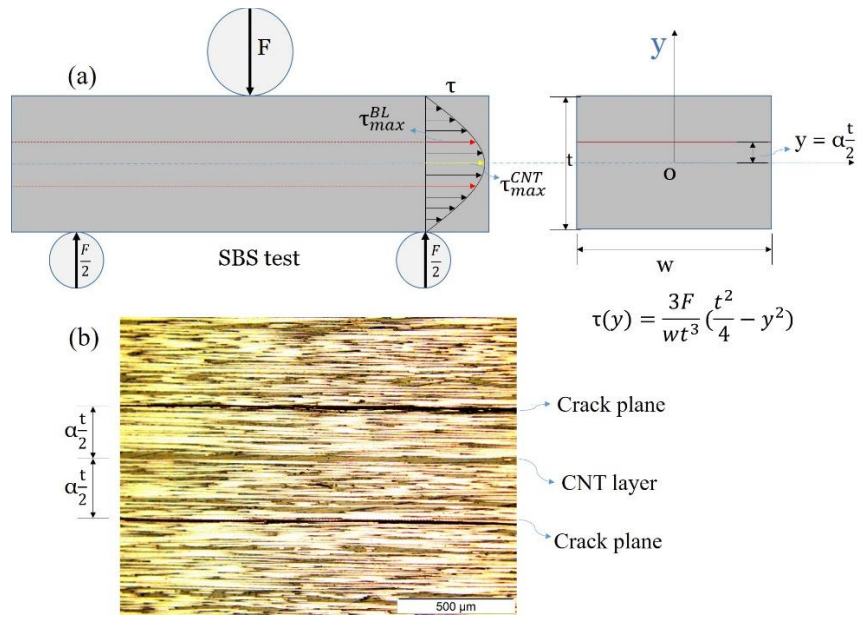

Fig. 7 (a) ILSS distribution of SBS specimen along the through-the-thickness direction, in which red and blue lines stand for crack plane and middle CNT layer, respectively; (b) cross-sectional images of tested SBS specimen interleaved with both CNT veils and resin film.

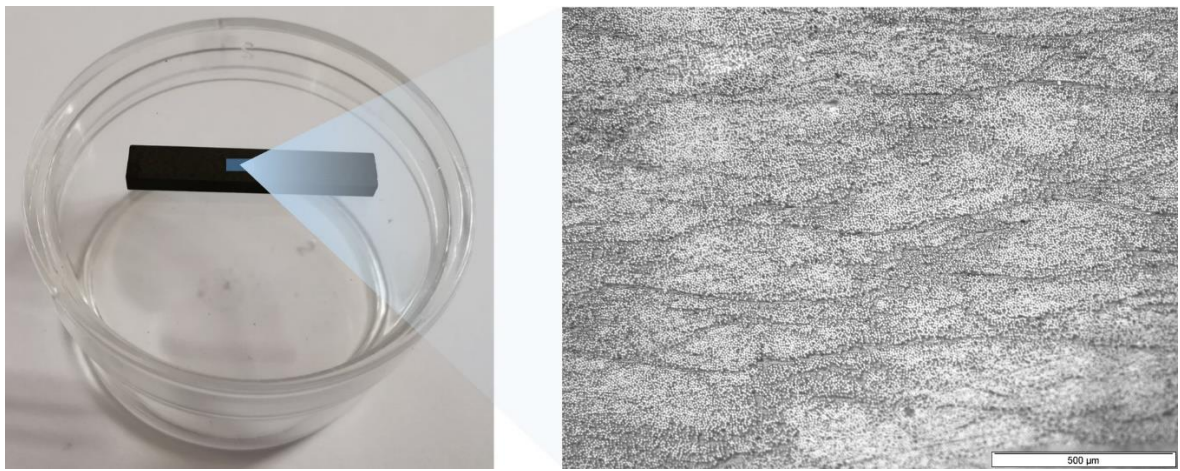

Fig. 8 AS4/8552 laminate interleaved with thin 8552 resin film showing no apparent increase of interlaminar thickness or resin rich regions in the middle ply compared with other interlaminar regions

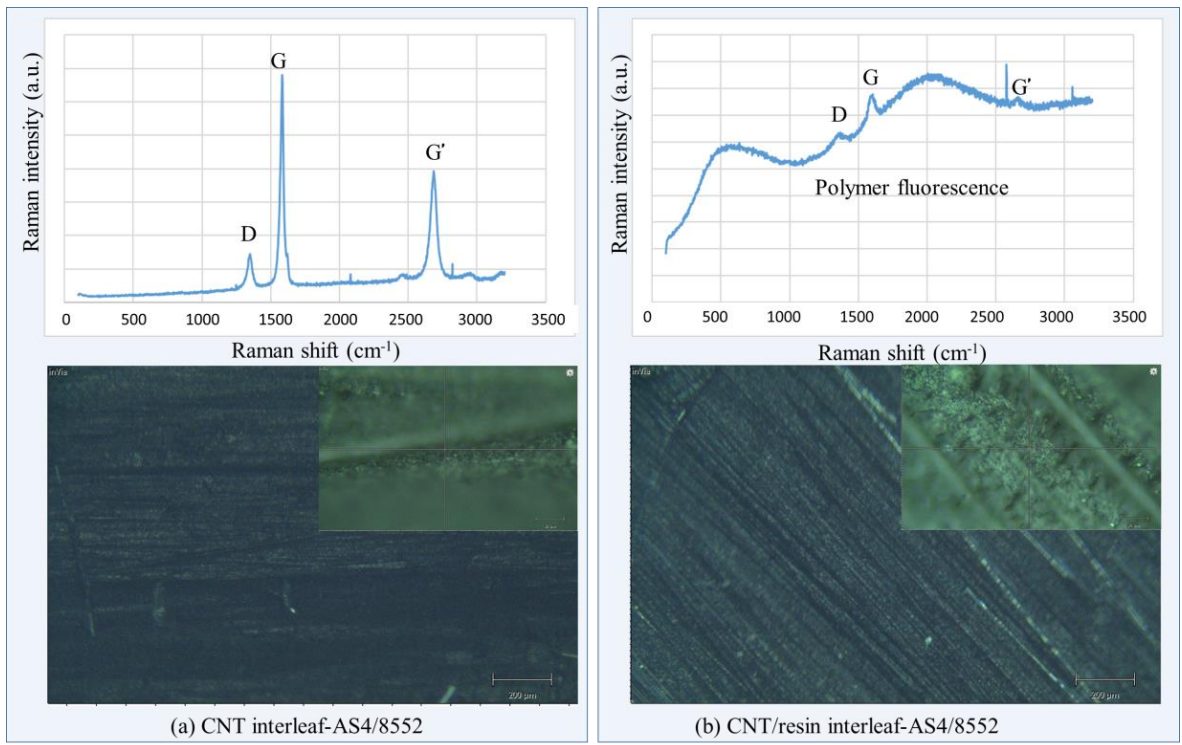

Fig. 9 Raman spectra of fracture surfaces in different specimens and corresponding optical micrographs. a) Strong features from the CNTs and no evidence of polymer, indicating no infiltration of the CNTs. b) CNT features superimposed on a fluorescent background and Raman features from infiltrated polymer

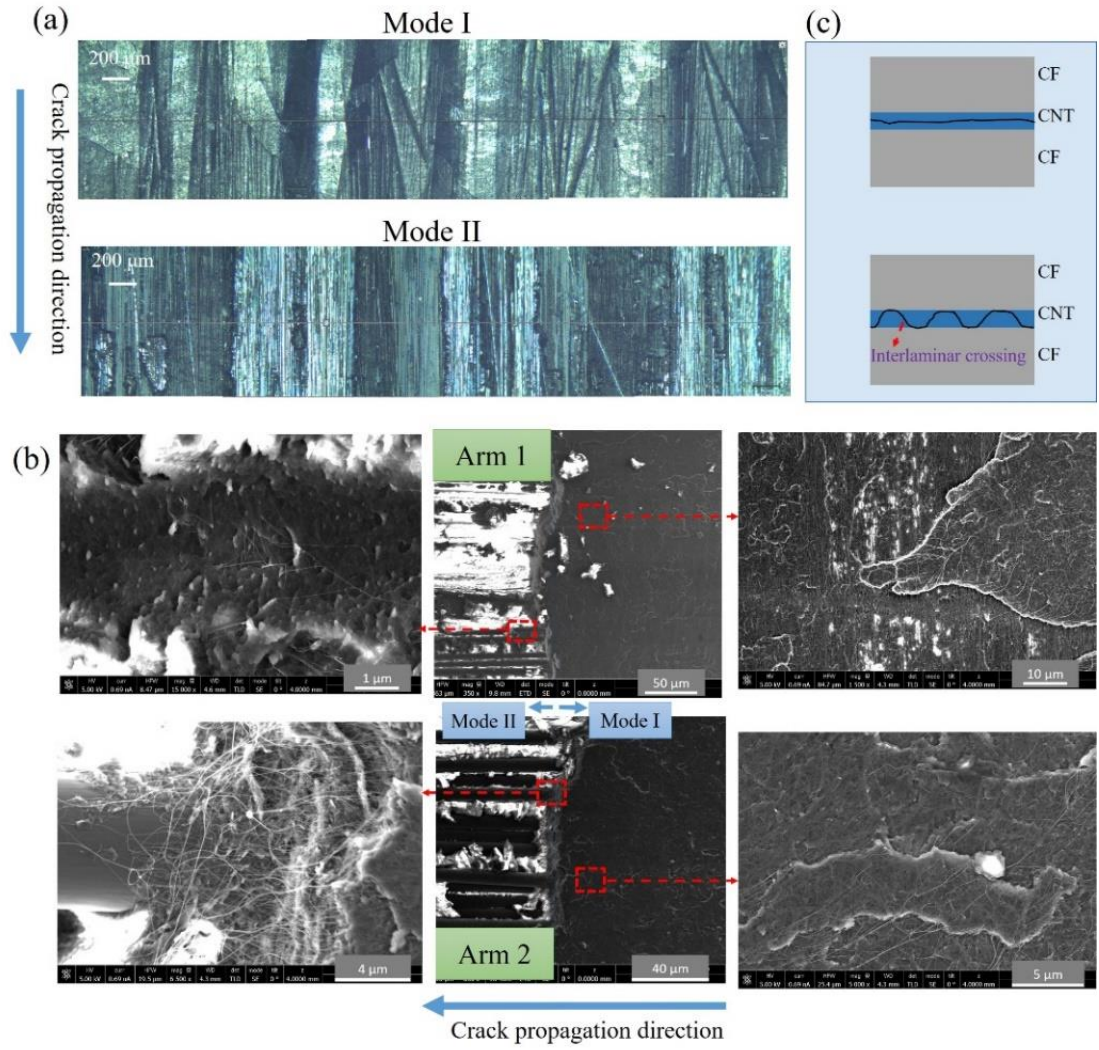

Fig. 10 (a) Mode I and Mode II fracture surfaces of CNT veil interleaved specimen (well-infiltrated); (b) Some details of transition zone from Mode I to Mode II and (c) different cracking behaviors under Mode I and Mode II loading condition.

Table 1 Comparison of Mode II toughening factor ($\eta_{II}$) by using different interleaf materials

| Interleaf material | Interleaf thickness/ ply thickness (μm/μm) | Mode II improvement (%) | $\eta_{II}$ | REF. |
|---|---|---|---|---|
| Nylon 6,6 nanofibrous mat | 40/210 | 61 | 3.2 | [50] |
| Nanotubes/epoxy film | 150/200 | 126.7 | 1.7 | [51] |
| As-prepared CNT BP/laminate | 20/200 | 34 | 3.4 | [52] |
| Cross-linked CNT BP/laminate | 20/200 | 74 | 7.4 | [52] |
| Nylon veils surface-loaded with silver nanowires | 53/125 | 227 | 5.4 | [53] |
| CNT veils | 10/180 | 88 | 15.8 | This work |

*Note: only research works using unidirectional CFRP composites were compared here to avoid other possible influences.*